\documentclass[pre,aps,twocolumn,superscriptaddress,showpacs]{revtex4}

\usepackage[dvips]{graphicx}
\usepackage{amssymb,amsfonts,amsmath}
\usepackage{color}
\usepackage{ulem}

\newcommand{\rv}{{\mathbf r}}

\newcommand{\e}{{\rm e}}
\newcommand{\J}{{\bf J}}

\newcommand{\pv}{{\bf p}}

\newcommand{\Xv}{{\bf X}}

\newcommand{\vel}{{\bf v}}

\newcommand{\rnt}{{\bf r}^N\!\!,t}
\newcommand{\kT}{k_{\rm B}T}

\newcommand{\GvH}{G_{\rm vH}}
\newcommand{\JvH}{{\bf J}_{\rm vH}}
\newcommand{\JvHf}{{\bf J}^{\rm f}_{\rm vH}}

\newcommand{\JvHfddft}{{\bf J}^{\rm f,DDFT}_{\rm vH}}

\newcommand{\kv}{{\bf k}}
\newcommand{\qv}{{\bf q}}

\begin{document}

\title{Nonequilibrium Ornstein-Zernike relation for Brownian many-body dynamics}

\author{Joseph M. Brader}
\affiliation{Department of Physics,
  University of Fribourg, CH-1700 Fribourg, Switzerland}
\email{joseph.brader@unifr.ch}

\author{Matthias Schmidt}
\affiliation{Theoretische Physik II, Physikalisches Institut, 
  Universit{\"a}t Bayreuth, D-95440 Bayreuth, Germany}

\date{Friday, 18 July 2013, J.\ Chem.\ Phys {\bf 139}, 104108 (2013),
http://dx.doi.org/10.1063/1.4820399}

\begin{abstract}
We derive a dynamic Ornstein-Zernike equation for classical fluids
undergoing overdamped Brownian motion and driven out of
equilibrium. Inhomogeneous two-time correlation functions are obtained
from functional differentiation of the one-body density and current
with respect to an appropriately chosen external field. Functional
calculus leads naturally to non-Markovian equations of motion for the
two-time correlators.  Memory functions are identified as functional
derivatives of a space- and time-nonlocal power dissipation
functional. We propose an excess (over ideal gas) dissipation
functional that both generates mode-coupling theory for the two-body
correlations and extends dynamical density functional theory for the
one-body fields, thus unifying the two approaches.
\end{abstract}

\maketitle

\maketitle

\section{Introduction}
In 1914 Leonard Ornstein and Frits Zernike developed a theory of
critical opalescence in which they proposed to separate the radial
distribution function, $g(r)$, into direct and indirect contributions
\cite{OrnsteinZernike1914}. The Ornstein-Zernike (OZ) relation, which
has since become a cornerstone of equilibrium liquid-state theory
\cite{Hansen06}, provides the mathematical expression of this
separation and defines the direct correlation function, $c(r)$, via
the integral relation
\begin{align}
  h(r_{13})=c(r_{13}) + \rho_b\!\int \!d\rv_2\,\, c(r_{12}) h(r_{23}),
\label{oz}
\end{align}
where $\rho_b$ is the bulk number density, $h(r)=g(r)-1$ is the total
correlation function, and $r_{ij}=|\rv_i-\rv_j|$.  The strength of
this approach lies in the fact that $c(r)$ usually has a simple
functional dependence on both the separation $r$ and on $\rho_b$, thus
facilitating the development of approximations.  By supplementing the
OZ relation with an appropriate closure relation between $h(r)$ and
$c(r)$, one arrives at a closed integral equation theory for the
equilibrium pair correlations, and hence for all thermodynamic
properties of the system \cite{Hansen06}.  Even simple, short-ranged
approximations to $c(r)$ can describe accurately the oscillatory
behaviour of $g(r)$, which arises from molecular packing effects, and
can capture the long-ranged decay of $g(r)$ near the critical point
\cite{FootnoteNOZ}, which causes the experimentally observed
turbidity.

Deeper insight into the OZ equation \eqref{oz}, as well as its natural
extension to inhomogeneous systems, is provided by modern density
functional theory (DFT) \cite{evans79}. Within DFT the direct
correlation function, $c(\rv_1,\rv_2)$, is defined as the second
functional derivative of the excess (over ideal gas) Helmholtz free
energy with respect to the density.  A second generating functional,
the grand potential, is obtained from Legendre transform of the
Helmholtz free energy and yields the density-density correlation
function,
$\rho(\rv_1)\rho(\rv_2)h(\rv_1,\rv_2)+\rho(\rv_1)\delta(\rv_1-\rv_2)$,
upon differentiation with respect to the external potential; here
$\rho(\rv)$ is the inhomogeneous one-body density distribution.  The
inhomogeneous OZ relation,
\begin{align}
h(\rv_1,\rv_3)=c(\rv_1,\rv_3)\, + \int \!d\rv_2\,\,
c(\rv_1,\rv_2)\rho(\rv_2)h(\rv_2,\rv_3),
\label{ozInhomogeneous}
\end{align}
which reduces to \eqref{oz} in the absence of an external potential,
then expresses the fact that the direct and density-density
correlation functions are (essentially) functional inverses of
  each other.  The OZ relation \eqref{ozInhomogeneous} thus plays the
role of a fundamental sum rule, distinct from hierarchies that relate
e.g.\ two-body functions to integrals over three-body functions
\cite{Hansen06}.  Higher-order correlation functions can be obtained
by further differentiation of the generating functionals
\cite{evans79} and are interrelated by higher-order OZ relations
\cite{LloydLee}.

The situation in nonequilibrium is quite different. No analogue of the
OZ relation is known that would determine {\it dynamic} correlation
functions \cite{zwanzig_book}.  Such a nonequilibrium Ornstein-Zernike
(NOZ) equation should fulfill at least three requirements: (i)~It
should determine the van Hove function, $\GvH(\rv_1,t_1,\rv_2,t_2)$,
which is the natural generalization of $g(\rv_1,\rv_2)$ to
time-dependent situations.  (ii)~It should be an equation on the
two-body level, distinct from the familiar $n$-body correlation
hierarchies \cite{Hansen06}.  (iii) Direct time correlation functions,
which depend on two points in spacetime, should occur, in analogy to
$c(\rv_1,\rv_2)$ in the static case. These conditions are not met by
simply adding a time argument to the functions appearing in
\eqref{ozInhomogeneous}, as has been suggested in the
literature~\cite{gan92}.

In this paper we propose a dynamical equation that satisfies all of
the above requirements.  The derivation is based on the dynamical
generalization of the well-known equilibrium concept of functional
differentiation as a means to generate $n$-point correlation
functions.  We first apply this method to dynamical density functional
theory \cite{marinibettolomarconi99}, and then supplement the
resulting approximate equation by a formally exact contribution that
involves direct time correlation functions. Furthermore, we show that
within the recently introduced power functional framework
\cite{power}, the direct time correlation functions can be identified
as second functional derivatives of the excess (over ideal gas) power
dissipation functional.  The NOZ equation may thus be closed via
approximation of the excess power dissipation functional, in the
spirit of equilibrium DFT.  Alternatively, in the spirit of integral
equation theory \cite{Hansen06}, one can postulate an additional
relation between the van Hove and the direct time correlation
functions. We show that mode-coupling theory can be viewed as
providing a closure of the latter type, where the memory function
plays the role of a direct time correlation function.

\section{Theory}
\subsection{Microscopic dynamics}\label{MicroscopicDynamics}
We describe the state of the system by a time-dependent distribution
function $\Psi(\rnt)$, which gives the probability density to find the
$N$ particles in the system at positions
$\rv^N\equiv\{\rv_1,\ldots\rv_N\}$ at time $t$. The particles interact
via an interparticle potential $U(\rv^N)$ and with their surrounding
via an external potential $V_{\rm ext}(\rv,t)$ and via a
non-conservative force field $\Xv(\rv,t)$.  The thermal agitation at
constant temperature $T$ is balanced by a velocity-dependent friction
force with force constant~$\gamma$. The resulting overdamped Brownian
dynamics can be described via the continuity equation for the
many-body distribution function,
\begin{align}
  \frac{\partial}{\partial t}\Psi(\rnt) = 
  -\sum_i \nabla_i\cdot \hat\vel_i(t) \Psi(\rnt).
  \label{smoluchowski}
\end{align}
Here the velocity operator of particle $i$ is defined as
\begin{align}\label{vel_op}
  \hat\vel_i(t) = \gamma^{-1}\Big[&-(\nabla_i U(\rv^N))
     - \kT \nabla_i \notag\\&
  -(\nabla_i V_{\rm ext}(\rv_i,t))
  +\Xv(\rv_i,t)  \Big],
\end{align}
where $k_{\rm B}$ is the Boltzmann constant, and $T$ is absolute
temperature.  Within this Smoluchowski picture the average of an
operator $\hat a(t)$ in configuration space is given by $a(t)=\langle
\hat a(t) \rangle = \int d\rv^N \hat a(t) \Psi(\rnt)$. 

\subsection{One-body averages}\label{OneBodyAverages}
For the present
study the one-body density and one-body current are of particular
importance and are described by the operators
\begin{align}\label{density_operator}
  \hat\rho(\rv,t) &= \sum_i \delta(\rv-\rv_i),  \\
  \label{current_operator}
  \hat\J(\rv,t) &= \sum_i \delta(\rv-\rv_i) \hat\vel_i(t).
\end{align}
For brevity we will henceforth use the shorthand notation
$\hat\rho(1) \equiv \hat\rho(\rv_1,t_1)$, and $\hat\J(1) \equiv
\hat\J(\rv_1,t_1)$ for spacetime points. The one-body density
and current are then given by $\rho(1)=\langle\hat\rho(1)\rangle$ and
$\J(1)=\langle\hat\J(1)\rangle$, respectively. The one-body velocity
is simply $\vel(1)=\J(1)/\rho(1)$. The local conservation of
particle number is expressed by the one-body continuity equation
\begin{align}
  \frac{\partial}{\partial t_1} \rho(1)
  &= -\nabla_1\cdot \J(1).
  \label{continuity1}
\end{align}

\subsection{Two-body correlation functions}\label{TwoBodyCorrelations}
On the two-body level, the molecular motion of a liquid is
commonly analyzed in terms of a two-time density-density correlation
function first introduced by Leon van Hove \cite{Hansen06}.  For
spatially and temporally inhomogeneous situations the van Hove
function is defined by
\begin{align}
  \GvH(1,2) & = \rho(1)^{-1}
  \langle \hat\rho(1)\hat\rho(2) \rangle,
  \label{EQGvH}  
\end{align} 
where the two-time average is taken with respect to the nonequilibrium
distribution at the earlier time $t_2$, together with the conditional
probability for finding the state at the later time $t_1$.

While the equilibrium relaxation dynamics of the system is
well-characterized by the van Hove function, the motion in the
presence of e.g.\ a time-dependent external potential or
non-conservative shear forces is better described by the
nonequilibrium two-body function
\begin{align}  
  \JvHf(1,2) &= \langle \hat\J(1)\hat\rho(2)\rangle,
  \label{EQJvH}  
\end{align}
which we will henceforth refer to as the (front) van Hove current.  We
adopt the convention $t_1\geq t_2$.  The analogue of
\eqref{continuity1} on the two-body level is given by
\begin{align}
  \frac{\partial}{\partial t_1} \rho(1) \GvH(1,2)
  &= -\nabla_1\cdot \JvHf(1,2),\label{continuity2}
\end{align}
which relates the vectorial van Hove current to the scalar van Hove
function.

\subsection{Static functional derivatives}\label{StaticDerivatives}
In order to connect the one-time level of description, provided by the
density, $\rho(1)$, and current, $\J(1)$, to the inhomogeneous
two-time van Hove current, we seek to express the latter as a
functional derivative of the former with respect to an
  appropriately chosen one-body field.  In equilibrium this procedure
is straightforward.  For example, the equilibrium density is given by
\begin{align}
\rho(\rv)={\rm Tr}_{\rm cl}
\hat{\rho}(\rv) \Psi_{\rm eq}(\rv^N),  
\end{align}
where ${\rm Tr}_{\rm cl}$ is the classical trace over phase space and
total particle number and the grand canonical probability density is
given by
\begin{align}
\Psi_{\rm eq}(\rv^N)=\Xi^{-1}e^{-\beta(H-\mu N)},
\label{eq_prob_density}
\end{align}
where $\Xi$ is the grand partition function, $\beta=(\kT)^{-1}$,
$H=\sum_i[{\bf p}_i^2/(2m)+V_{\rm ext}(\rv_i)]+U(\rv^N)$ is the
Hamiltonian, ${\bf p}_i$ is the momentum of particle $i$, $m$ is the
particle mass, and $\mu$ is the chemical potential \cite{Hansen06}.
Functional differentiation of the density with respect to its
conjugate field, the external potential, generates the density-density
correlation function
\begin{align}
\frac{\delta \rho(\rv)}{\delta \beta V_{\rm ext}(\rv')} \Big|_{\rm eq}
=\langle \hat{\rho}(\rv)\hat{\rho}(\rv') \rangle 
-\rho(\rv)\rho(\rv')\,, 
\label{eq_derivative}
\end{align} 
Here we use that for fields $u(\rv)$ that depend only on space $\delta
u(\rv)/\delta u(\rv')=\delta(\rv-\rv')$.

\subsection{Dynamic functional derivatives}\label{DynamicalDerivatives}
Out of equilibrium there is no standard procedure for generating, in
the spirit of \eqref{eq_derivative}, inhomogeneous two-time
correlation functions.  We thus seek to express microscopic two-time
correlators as functional derivatives of one-body fields.  Consider
the Smoluchowski equation \eqref{smoluchowski} in the form
\begin{align}
  \frac{\partial}{\partial t}\Psi(\rnt) = 
  \hat\Omega(t) \Psi(\rnt),
  \label{smoluchowski_operatorform}
\end{align}
where the (Smoluchowski) time  evolution operator, given by
\begin{align}\label{EQomegaDefinition}
    \hat{\Omega}(t) &= -\sum_i \nabla_i\cdot\hat\vel_i(t),
\end{align}
allows to write the formal solution of
\eqref{smoluchowski_operatorform} as
\begin{align}
  \Psi(\rnt) = \e_+^{\int_{t_0}^t \!ds\, \hat{\Omega}(s)}\Psi(\rnt_0),
\label{smol_propagator}  
\end{align}
where $t_0$ is an initial time and $\e_+$ indicates a time-ordered
exponential (see e.g.~\cite{braderMCT2012}), which is defined via the
power series
\begin{align}
  \e_+^{\int_{t_0}^tds\hat\Omega(s)}&=
  1+\int_{t_0}^tds\hat\Omega(s)+
  \int_{t_0}^tds_1\int_{t_0}^{s_1}ds_2\hat\Omega(s_1)\hat\Omega(s_2)\notag\\
  &\hspace{-5mm}+\int_{t_0}^tds_1\int_{t_0}^{s_1}ds_2\int_{t_0}^{s_2}ds_3
  \hat\Omega(s_1)\hat\Omega(s_2)\hat\Omega(s_3)+\ldots
\end{align}
The time-ordered exponential in \eqref{smol_propagator} acts as a
propagator and will play a role analagous to that of the Boltzmann
factor in the equilibrium distribution \eqref{eq_prob_density}.

In order to calculate the desired functional derivatives we will use
the general functional identity $\delta u(\rv,t)/\delta u(\rv',t') =
\delta(\rv-\rv')\delta(t-t')$, where $u(\rv,t)$ is an arbitrary
function, and furthermore the chain rule for time-ordered
exponentials,
\begin{align}\label{propagator_deriv}
\frac{\delta}{\delta u(\rv,t)}\e_+^{\int_{t_1}^{t_2} \!ds \,\hat\Omega(s)}=
\int_{t_1}^{t_2}\!\!ds\,
  \e_+^{\int_{s}^{t_2} \!ds'\, \hat\Omega(s')}
  \frac{\delta \hat\Omega(s)}{\delta u(\rv,t)}
  \, \e_+^{\int_{t_1}^{s} \!ds' \hat\Omega(s')}.
\end{align}
Observing the general definition of the two-time correlation 
between two operators $\hat a(1)$ and $\hat b(2)$,
\begin{align}
\hspace{-2.1mm}  \langle \hat a(1) \hat b(2)\rangle = 
  \int d\rv^N 
  \hat a(1)
  \e_+^{\int_{t_2}^{t_1}ds\hat\Omega(s)}
  \hat b(2)
  \e_+^{\int_{t_0}^{t_2}ds\hat\Omega(s)}
  \psi(\rnt_0),
  \label{EQdefintionTwoTimeCorrelator}
\end{align}
and using
\eqref{propagator_deriv} it is straightforward to show that the
following functional derivative relations hold
\begin{align}
  &\frac{\delta \J(1)}{\delta \beta\mathcal{V}(2)} =
  I(1,2)
+ \frac{\partial}{\partial t_2}
   \JvHf(1,2),\label{EQJbyX}\\
  &\frac{\delta \rho(1)}{\delta \beta\mathcal{V}(2)} =
  \rho(1)\frac{\partial}{\partial t_2}\GvH(1,2)
  \label{EQrhoByX}.
\end{align}
where causality requires $t_2\leq t_1$.  The functional derivatives
are built with respect to the function
\begin{align}
\mathcal{V}(2)\equiv \int_{t_0}^{t_2}\!\!dt_2'\,D_0\nabla^2_2 \, V_{\rm ext}(2'), 
\end{align}
where we employ the notation $V_{\rm ext}(2')=V_{\rm
  ext}(\rv_2,t'_2)$.  The function $\mathcal{V}(2)$ has the same
physical dimension as the external potential, but rather measures the
accumulated change in potential arising from the action of the
diffusion operator. The instantaneous contribution to \eqref{EQJbyX}
is given by $I(1,2)=-\gamma^{-1}\rho(1) \delta\nabla V_{\rm
  ext}(1)/\delta \beta \mathcal{V}(2)$; explicit evaluation of the
functional derivative will not be required for the following
development.

The consistency of our formalism with the
equilibrium density functional approach can be demonstrated by assuming the system was in equilibrium 
for all times and integrating the dynamic functional derivative \eqref{EQrhoByX} over the 
entire history
\begin{align}
  \int_{-\infty}^{t_2}\!\!dt_2'\,\,\frac{\delta \rho(1)}{\delta \beta\mathcal{V}(2')}&=
  \!\int_{-\infty}^{t_2}\!\!dt_2'\,\,
  \rho(1)\frac{\partial}{\partial t'_2}\GvH(1,2'),
  \\  
&=\langle\hat{\rho}(\rv_1)\hat{\rho}(\rv_2) \rangle 
- \rho(\rv_1)\rho(\rv_2)
\label{recovered}
\\
&
= \frac{\delta \rho(\rv_1)}{\delta \beta V_{\rm ext}(\rv_2)} \Big|_{\rm eq},  
\label{noneq-eq}
\end{align}
where we make the (reasonable) assumption that density fluctuations
become decorrelated at sufficiently long times.  Note that in this
dynamical calculation the second term in \eqref{recovered} arises from
the lower integration limit, whereas in the standard Gibbs ensemble
calculation \eqref{eq_derivative} it is generated by the normalization
of the probability distribution.

\subsection{DDFT Approximation}
\label{DDFTapproximation} 
We next seek to apply the mathematical framework developed above to
generate equations of motion for the two-time correlation functions.
This requires explicit expressions for the one-body averages which can
be differentiated with respect to the external fields.  The simplest
theory for the microscopic one-body current of interacting particles
is the dynamical density functional theory (DDFT)
\cite{marinibettolomarconi99}, where the current,
\begin{align}
  \J_{\rm DDFT}(1) = \frac{\rho(1)}{\gamma}\left(
  -\nabla \frac{\delta F[\rho]}{\delta \rho(1)} - \nabla V_{\rm ext}(1) +\Xv(1)
  \right), 
  \label{ddftCurrent}
\end{align}
expresses a time-local balance between the viscous friction,
$\gamma\vel(1)$, external forces, forces due to thermal motion and
interparticle interactions, the latter two contributions generated by
the intrinsic Helmholtz free energy functional $F[\rho]$.  When
combined with the one-body continuity equation \eqref{continuity1}, a
closed drift-diffusion equation for $\rho(1)$ follows.

Using \eqref{ddftCurrent} to calculate the functional derivative
$\delta \J(1)/\delta \beta \mathcal{V}(3)$, employing the functional
chain rule, and the relations \eqref{EQJbyX} and \eqref{EQrhoByX}
generates a DDFT approximation to the van Hove current,
\begin{align}\label{TwoBodyCurrentEquation} 
  \J^{\rm f,DDFT}_{\rm vH}(1,3)&= \J(1)\GvH(1,3) -D_0\rho(1) \nabla_1 
  \Big(\GvH(1,3)
  \notag\\&\hspace*{-0.85cm}
  -\int d\rv_2 c(1,2_1) \rho(2_1) \left(\GvH(2_1,3)\!  
  -\! \rho(3_{-\infty})\right)\Big),
\end{align} 
where $\rho(3_{-\infty})\equiv \rho(\rv_3,-\infty)$ and a contribution
$\nabla_1 \rho(3_{-\infty})$ vanishes.  The argument $2_1$ in
\eqref{TwoBodyCurrentEquation} indicates position $\rv_2$ and time
$t_1$; the direct correlation function is hence evaluated at distinct
values of the spatial arguments at the same time, $c(1,2_1)\equiv
c(\rv_1,\rv_2,t_1)$, and ${\bf v}(1)$ is given here by $\J_{\rm
  DDFT}(1)/\rho(1)$.  Here the equilibrium direct correlation function
is the second functional derivative of the excess (over ideal gas)
part of the intrinsic Helmholtz free energy, $c(\rv_1,\rv_2)\!=\!-
\delta^2 \beta F^{\rm exc}[\,\rho]/\delta\rho(\rv_1)\delta\rho(\rv_2)$
\cite{evans79}. In obtaining \eqref{TwoBodyCurrentEquation} we
have made the assumption that two-body correlations factorize for
widely separated time arguments, i.e.  $\langle
\hat{\rho}(\rv,t)\hat{\rho}(\rv',-\infty)
\rangle=\rho(\rv,t)\rho(\rv',-\infty)$, which holds in the absence of
an ideal glass transition.  The three distinct contributions to
\eqref{TwoBodyCurrentEquation} represent a transport term, ideal
decay, and an adiabatic integral term due to interparticle
interactions.

Equation \eqref{TwoBodyCurrentEquation} is the natural extension of
the DDFT approximation for the one-body current \eqref{ddftCurrent} to
the two-body level. Substitution of \eqref{TwoBodyCurrentEquation}
into the two-body continuity equation \eqref{continuity2} yields a
closed equation for the van Hove function which is local in time, due
to the adiabatic assumption underlying \eqref{ddftCurrent}, but
nonlocal in space; this is the DDFT approximation to the NOZ equation
we seek.  External forces do not appear in
\eqref{TwoBodyCurrentEquation} explicitly, but enter implicitly via
the one-body density and current obtained by solving
\eqref{continuity1} with \eqref{ddftCurrent}.  The fact that equation
\eqref{TwoBodyCurrentEquation} is closed is a direct consequence of
the adiabatic assumption that thermodynamic driving forces can be
generated from an equilibrium free energy functional. As we will
discuss below, this is equivalent to neglecting the contribution of
interparticle interactions to the the power dissipation in the
dynamical generating (power) functional.

Within the same DDFT approximation considered here Archer {\it et al.}
\cite{archer07dtplhopkins10dtpl_1,archer07dtplhopkins10dtpl_2} have
proposed a dynamic test-particle method for calculating the
equilibrium van Hove function.  This alternative approach focuses on
the simultaneous relaxation of both a tagged particle density (from a
delta-function initial state) and the one-body density distribution of
the remaining particles (from initial state $\rho(r,0)=\rho_b g(r)$).
In general, this one-body route to the van Hove function will produce
results which differ from those generated by Equations
\eqref{continuity2} and \eqref{TwoBodyCurrentEquation}. In the special
case that the Helmholtz free energy is approximated by a quadratic
density expansion (the Ramakrishnan-Yussouff (RY) approximation
\cite{rama}) the test-particle current becomes identical to
\eqref{TwoBodyCurrentEquation}. The RY functional thus exhibits
test-particle self consistency within the DDFT approximation.

\subsection{Equal-time equilibrium correlations}
In the special case of equilibrium at all times, $\J(1)\!=\!0$, the
equal-time limit, $t_1=t_3$, of \eqref{TwoBodyCurrentEquation} yields
\begin{align}\label{EqualTimeLimit} 
&\;\;\J^{\rm f,DDFT}_{\rm vH}(1,3_1)=  -D_0\rho(1) \nabla_1 
\Big[\delta(\rv_1-\rv_3) \\  &
+\rho(3_1)\Big( h(1,3_1)\!-\! c(1,3_1)\!-\!\!\! \int d\rv_2 c(1,2_1)\rho(2_1)h(2_1,3_1)
\Big)\Big],\notag
\end{align}
where we have used the equal time limit of the van Hove function,
$G_{\rm vH}(1,3_1)=\rho(3_1)(h(1,3_1)+1)+\rho(1)\delta(\rv_1-\rv_3)$.
The short-time relaxation of the van Hove function is determined by
the highly localised, delta-function initial condition of the self
part, such that the term in square brackets in \eqref{EqualTimeLimit}
is identically zero. The dynamic functional derivative approach
  to two-time correlation functions thus provides an alternative
  derivation of the inhomogeneous OZ equation~\eqref{ozInhomogeneous}.

\subsection{Homogeneous system without external forces}
In the homogeneous limit with no external forces equation
\eqref{TwoBodyCurrentEquation} reduces to
\begin{align}
  \frac{\partial}{\partial t}F(k,t) 
  + \Gamma(k)F(k,t)=0, 
\label{MCT_DDFT}
\end{align} 
where the intermediate scattering function, $F(k,t)$, is the
three-dimensional spatial Fourier transform of the translationally
invariant equilibrium van Hove function \cite{Hansen06}.  The
time-independent `initial decay rate' is given by
$\Gamma(k)\!=\!D_0\,k^2/S(k)$, where
$S(k)\!=\!1/(1\!-\rho_b\tilde{c}(k))$ is the static structure factor;
here the tilde indicates the spatial Fourier transform.
Equation~\eqref{MCT_DDFT} has the solution
\begin{align}
  F(k,t) = e^{-\Gamma(k)t}.
  \label{deGennes} 
\end{align}   
The effective diffusion constant, $D_0/S(k)$, is strongly reduced for
wavenumbers around the main peak of $S(k)$, relative to the bare
diffusion constant.  This well-known `de Gennes narrowing'
\cite{deGennesNarrowing} has its origins in the strong spatial
correlations at wavelengths corresponding to the local
nearest-neighbour cage around any given particle.

\subsection{Homogeneous system under shear}
When applied to a spatially homogeneous systems under steady shear of
rate $\dot\gamma_s$, with flow in $x$-direction and shear-gradient in
$y$-direction, Eq.\ \eqref{ddftCurrent} yields $\gamma\J_{\rm
  DDFT}(1)/\rho_b=\Xv(1)\equiv \dot{\gamma}_s y_1\hat{\bf e}_x$.
Solution of Eqs.\ \eqref{continuity2} and
\eqref{TwoBodyCurrentEquation} for this one-body current is
straightforward (using, e.g.\ the method of characteristics) and
yields
\begin{align}
  F(\kv,t) = \exp\left(\!-\frac{\kv\kv 
    \!:\!{\bf D}(t\,;\dot{\gamma}_s)}{S(k(t))}\right),
  \label{deGennes_flow} 
\end{align} 
in which the wavevector dyadic $\kv\kv$ is fully contracted with the
time-dependent diffusion tensor, given by
\begin{align}
{\bf D}(t\,;\dot{\gamma}_s)=
\begin{pmatrix} 
t + \frac{\dot{\gamma}_s^2 t^3}{3} & \dot{\gamma}_st^2 & 0 
\\ 
0 & t & 0 
\\ 
0 & 0 & t  
\end{pmatrix}, 
\end{align}
and where the shear-advected wavevector is given by
$\kv(t)\!=\!(k_x,k_y\!+\dot{\gamma}_sk_xt,k_z)$.  Equation
\eqref{deGennes_flow} extends \eqref{deGennes} to steadily sheared
states and captures the enhanced diffusion in flow direction, termed
`Taylor dispersion' \cite{taylordispersion}, which arises from the
coupling of Brownian and affine motion.  Equation
\eqref{TwoBodyCurrentEquation} thus treats systems with non-zero
density by supplementing the exact low density limit (equation
\eqref{deGennes_flow} with $S(k)\!=\!1$) with an approximate,
wavevector dependent diffusion tensor. 
The approximation \eqref{deGennes_flow} is on a similar level to the 
fluctuating diffusion equation approach of Ronis \cite{ronis} (for a review 
of alternative approaches to calculating the distorted structure factor 
see \cite{brader10review}).

\subsection{Beyond DDFT}\label{beyondDDFT}
In contrast to the DDFT approximation \eqref{TwoBodyCurrentEquation},
the exact expression for $\JvHf(1,2)$ should include the physics of
structural relaxation, via a dependence on the history of both one-
and two-body correlation functions.  Introduction of vectorial and
tensorial direct time correlation functions, denoted by ${\bf m}(1,2)$
and ${\sf M}(1,2)$, respectively, enables formulation of a general
equation of motion.  Splitting the full van Hove current into the DDFT
contribution and an irreducible part, $\JvHf=\JvHfddft+\JvH^{\rm
  irr}$, we identify the most general non-Markovian form that involves
only one- and two-body functions which generate a vector field from
spacetime convolutions of the van Hove function and van Hove current,
namely
\begin{align}
  \label{TwoBodyCurrentEquationGeneral}
  &\JvH^{\rm irr}(1,3)=\JvH^{\rm irr}(1,3_{-\infty})
  -\rho(1)\!\int_{-\infty}^{\,t_3} \!\!\!dt'_3\, \nabla_3\cdot{\sf M}(1,3')\rho(3') 
  \notag\\  & \quad
    +\rho(1)\! \int \!d2\, \Big[{\sf M}(1,2)\cdot(\JvHf(2,3)-\J(2)\rho(3_{-\infty}))\notag\\
    & \quad
    \hspace*{1.4cm}+{\bf m}(1,2)\rho(2)(\GvH(2,3)-\rho(3_{-\infty}))\Big],
\end{align}
The $2$-integral in \eqref{TwoBodyCurrentEquationGeneral} runs over a
spacetime slab from the earlier time $t_3$ to later time $t_1$,
consistent with causality.  Unlike the approximate DDFT expression
\eqref{TwoBodyCurrentEquation}, the exact NOZ equation is not closed
and serves to define the direct time correlation functions ${\bf
  m}(1,2)$ and ${\sf M}(1,2)$, in analogy to the equilibrium OZ
equation \eqref{ozInhomogeneous}, which defines the static direct
correlation function $c(\rv_1,\rv_2)$. Equation
\eqref{TwoBodyCurrentEquationGeneral}, when combined with
\eqref{TwoBodyCurrentEquation}, provides a natural generalization of
the equilibrium OZ equation to nonequilibrium systems undergoing
Brownian dynamics and enables structural relaxation to be incorporated
via the time direct correlation functions.  Although the continuity
equation \eqref{continuity2} can be used to eliminate $\JvHf(1,3)$
from \eqref{TwoBodyCurrentEquationGeneral} in favour of $\GvH(1,3)$,
closure still requires that \eqref{TwoBodyCurrentEquationGeneral} be
supplemented by two independent equations that relate ${\bf m}(1,2)$
and ${\sf M}(1,2)$ to the van Hove function and its current.  This can
be achieved either by postulating closure relations (as is done in
equilibrium via e.g.\ the Percus-Yevick or hyper-netted-chain
approximations \cite{Hansen06}), or by exploiting the power functional
formalism \cite{power}, as outlined below.

An approximation of particular significance is obtained by setting
${\bf m}(1,2)\!=\!0$, neglecting the second direct (without spatial
convolution) term, and simplifying the tensorial structure of the
remaining direct time correlation function, ${\sf
  M}(1,2)\!=\!M(1,2){\bf 1}$, where $M(1,2)$ is a scalar function and
$\bf 1$ the unit matrix.  For homogeneous equilibrium states the
general equation \eqref{TwoBodyCurrentEquationGeneral} then reduces to
\begin{align}
  \frac{\partial}{\partial t}F(k,t) + \Gamma(k)F(k,t)
  - \!\int_0^{t}\!dt' \tilde{M}(k,t\!-\!t')\dot{F}(k,t')=0, 
\label{MCT_equationofmotion}
\end{align}
which is a non-Markovian equation
for the intermediate scattering function, identical to that employed
in mode-coupling theory (MCT). The standard `idealized' mode-coupling
theory \cite{goetze_book} is obtained by setting
$\tilde{M}(k,t)\!=\!\tilde{M}_{\rm MCT}(k,t)$, where the friction kernel is 
given by
\begin{align}
\tilde{M}_{\rm MCT}(k,t)=-\frac{\rho_b \Gamma(k)}{16\pi^3}
\!\int \!d\qv \,\,V(\kv\,,\qv)\,F(q,t)F(|\kv-\qv|,t)
\label{MCT_closure}
\end{align}
contains the vertex function 
\begin{align}
\!\!V(\kv\,,\qv)=\frac{S(k) S(q) S(p)}{k^4}\Big(
\kv\cdot\qv \,\tilde{c}(q) + \kv\cdot\pv \,\tilde{c}(p)
\Big)^2,
\label{vertex}
\end{align}
where $\pv=\kv-\qv$.  The equations of MCT,
\eqref{MCT_equationofmotion}--\eqref{vertex}, capture slow structural
relaxation, absent from the simple DDFT approximation
\eqref{deGennes}, and predict dynamical arrest in dense and/or
strongly attractive systems \cite{goetze_book}.  The MCT closure
\eqref{MCT_closure} is local in time but nonlocal in space. Relaxation
of the restriction ${\bf m}(1,2)=0$, generates an additional term
within the time integral in \eqref{MCT_equationofmotion}, which is
linear in the intermediate scattering function. Remarkably, this
extension, which emerges naturally within the NOZ approach, is
consistent with the `extended MCT' of Gotze and Sjogren
\cite{sjoegren_hopping}, in which an additional relaxation process was
introduced to describe relaxation processes (`hopping') in glassy
states.

\subsection{Connection to power functional theory}\label{TheConnection}
The NOZ approach developed in this work gains further significance
when viewed in the context of the recently developed power functional
theory \cite{power}, which is an extension of classical density
functional theory to nonequilibrium.  Within this framework,
minimization of a dynamic (free power) functional yields a general and
exact equation of motion for the one-body current,
\begin{align}
 \J(1) &=
 \J_{\rm DDFT}(1) - \frac{\rho(1)}{\gamma}
  \frac{\delta P_{t_1}^{\rm exc}[\rho,\J]}{\delta \J(1)},
  \label{EQcurrentPFT}
\end{align}
where $\J_{\rm DDFT}(1)$ is defined via \eqref{ddftCurrent} and
obtained by differentiation of the ideal gas contribution to the power
dissipation, $P_t^{\rm id}[\rho,\J]=\int d\rv \gamma
\J(\rv,t)^2/(2\rho(\rv,t))$ with respect to the one-body current.  The
excess (over ideal gas) power dissipation, $P_t^{\rm exc}[\rho,\J]$,
is a functional of the history of $\rho(1)$ and $\J(1)$ prior to
time~$t$ and accounts for dissipation induced by particle-particle
interactions.  Differentiating the exact Euler-Lagrange equation
\eqref{EQcurrentPFT} with respect to $\beta\mathcal{V}(3)$ (following
the same procedure used to obtain \eqref{TwoBodyCurrentEquation}) and
comparing the result to the general
form~\eqref{TwoBodyCurrentEquationGeneral} yields the identification
of the direct time correlation functions with second functional
derivatives of the excess power dissipation via
\begin{align}
  {\bf m}(1,2) &= -\gamma^{-1}\frac{\delta}{\delta \rho(2)} 
  \frac{\delta P_{t_1}^{\rm exc}[\rho,\J]}{\delta \J(1)},
  \label{EQdefinitionDTCFvector}\\
  {\sf M}(1,2)^{\sf T} &= -\gamma^{-1}\frac{\delta}{\delta \J(2)} 
  \frac{\delta P_{t_1}^{\rm exc}[\rho,\J]}{\delta \J(1)},
  \label{EQdefinitionDTCFmatrix}
\end{align}
where the superscript $\sf T$ indicates the transpose.  Equations
\eqref{TwoBodyCurrentEquationGeneral} and
\eqref{EQcurrentPFT}--\eqref{EQdefinitionDTCFmatrix} imply that
approximating a single mathematical object, the excess power
dissipation functional, is sufficient to generate a closed and fully
consistent set of equations for the dynamics of both the one- and
two-body correlation functions.  The DDFT approximation, leading to
Eqs.\ \eqref{ddftCurrent} and \eqref{TwoBodyCurrentEquation}, is
obtained by setting $P_t^{\rm exc}[\rho,\J]=0$.

A natural way to go beyond DDFT is to approximate $P_t^{\rm
  exc}[\rho,\J]$ by a truncated (functional) Taylor expansion in the
one-body current about the equilibrium state.  To leading (quadratic)
order this yields
\begin{align}
P_t^{\rm exc}[\rho,\J]\, =-\frac{\gamma}{2}\!
\int\! d\rv_1 \!\int \!d2\;\; \J(1)\cdot
{ {\sf M}(1,2) \Big|}_{_{\J=0}} \!\!\!\!\!\!\!\!\cdot\J(2),
\label{quadratic_approximation}
\end{align}   
Making the approximation
$\left.{\sf M}(1,2)\right|_{\J=0}\!\approx M_{\rm MCT}(1\!-\!2){\bf
  1}$, which on the two-body level yields the MCT equation, generates
on the one-body level a closed non-Markovian equation of motion for
the current and density, via \eqref{continuity1} and
\eqref{EQcurrentPFT}.  This equation of motion contains information
about slow structural relaxation, on the level of idealized MCT, and
thus provides a useful tool to study situations for which standard
DDFT fails, such as e.g.\ the sedimentation of colloidal gels
\cite{brambilla2011}.

\section{Concluding remarks}\label{conclusions}
To summarize, we have developed a nonequilibrium Ornstein-Zernike
approach to the two-time correlation functions of interacting Brownian
particles.  The most fundamental equations emerging from our treatment
are \eqref{TwoBodyCurrentEquation} and
\eqref{TwoBodyCurrentEquationGeneral} for the vectorial van Hove
current. 
When supplemented by the two-body continuity
equation \eqref{continuity2}, these expressions provide a means to
calculate the two-time dynamical correlation functions under the
influence of arbitrary external forces.  Approximate closures, of
which MCT is a specific nontrivial example, generate, in general,
non-Markovian equations of motion for the van Hove function
\eqref{EQGvH} and the van Hove current \eqref{EQJvH}.

As part of our development of the general nonequilibrium theory we
have derived a DDFT approximation for the van Hove current
\eqref{TwoBodyCurrentEquation}. This expression, which to the best of
our knowledge has not appeared previously in the literature, provides
much confidence in our general approach and strongly supports our
identification of the microscopically defined two-time correlation
functions, namely the van Hove function \eqref{EQGvH} and van Hove
current \eqref{EQJvH}, as functional derivatives of the one-body
fields, via \eqref{EQJbyX} and \eqref{EQrhoByX}.  The DDFT
approximation to the van Hove current \eqref{TwoBodyCurrentEquation}
predicts de Gennes narrowing of the intermediate scattering function
for homogeneous systems and derives the inhomogeneous equilibrium OZ
relation \eqref{ozInhomogeneous} in the equilibrium limit, consistent
with the underlying adiabatic approximation.

One can view the general equations of motion
\eqref{TwoBodyCurrentEquation} and
\eqref{TwoBodyCurrentEquationGeneral} as the basis for the
construction of approximate nonequilibrium integral equation theories.
However, an alternative, and potentially more illuminating approach to
a closed two-time theory is provided by the power functional formalism
\cite{power}.  Within this framework, non-adiabatic contributions to
the one-body equation of motion \eqref{EQcurrentPFT} and the memory
functions entering the two-time equations
\eqref{EQdefinitionDTCFvector} and \eqref{EQdefinitionDTCFmatrix}, are
related to first and second derivatives, respectively, of the excess
power dissipation, as a single generating functional.  In analogy with
equilibrium DFT, for which the direct correlation function,
$c(\rv_1,\rv_2)$, is generated from a free energy functional, the
nonequilibrium time-direct correlation functions (memory functions)
responsible for non-Markovian dynamics are generated by the excess
power dissipation functional.  As the same excess power dissipation
functional generates the dynamics of the one-body fields, via the
Euler-Lagrange equation \eqref{EQcurrentPFT}, the power functional
approach can be seen to provide a unified variational framework for
the calculation of one- and two-body dynamical correlation functions.
Further functional differentiation of
\eqref{TwoBodyCurrentEquationGeneral} with respect to external forces
generates higher-order NOZ relations involving, for example, three-
and four-body correlations.  Despite their complexity, these
expressions may be of use, perhaps in a simplified form or in special
limits, for analyzing dynamical heterogeneities in equilibrium or in
metastable, arrested states~\cite{binderkob}.

By expressing the MCT within the framework of functional
differentiation, it is straightforward to identify a candidate
approximation, \eqref{quadratic_approximation}, to the excess power
dissipation functional, which can then be equally well applied, via
\eqref{EQcurrentPFT}, on the one-body level.  This opens up the
possibility of exploiting approximations developed on the two-body
level (e.g. MCT) for one-body variational calculations, thus putting
within reach the systematic investigation of many interestesing
problems for which the one-body dynamics may be significantly
influenced by slow structural relaxation (e.g. the sedimentation of
gels \cite{brambilla2011}).  When applied to inhomogeneous driven
systems the approximate excess dissipation functional
\eqref{quadratic_approximation}, together with \eqref{continuity2} and
\eqref{TwoBodyCurrentEquationGeneral}, provides a natural extension of
equilibrium MCT.  Moreover, application of
\eqref{quadratic_approximation} in a dynamic test particle calculation
of the type developed by Archer {\it et al.}
\cite{archer07dtplhopkins10dtpl_1,archer07dtplhopkins10dtpl_2} may
provide results for the intermediate scattering function which are
superior to the standard MCT, as the calculation is performed at the
one-body level.  Research along all these lines is in progress.

\subsection*{Acknowledgements}
We thank A. Fortini and D. de las Heras for useful discussions.

\end{document}